\documentclass[useAMS]{mn2e}
\usepackage{times}
\usepackage{rotating}
\long\def\symbolfootnote[#1]#2{\begingroup%
\def\thefootnote{\fnsymbol{footnote}}\footnote[#1]{#2}\endgroup}
\newcommand{\gae}{\lower 2pt \hbox{$\, \buildrel {\scriptstyle >}\over {\scriptstyle
\sim}\,$}}
\newcommand{\lae}{\lower 2pt \hbox{$\, \buildrel {\scriptstyle <}\over {\scriptstyle
\sim}\,$}}

\begin{document}

\title[GRB high energy photons detected by Fermi]
{On the generation of high energy photons detected by the Fermi 
Satellite from gamma-ray bursts}

\author[Kumar \& Barniol Duran]{P. Kumar$^{1}$\thanks
{E-mail: pk@astro.as.utexas.edu, rbarniol@physics.utexas.edu}
and R. Barniol Duran$^{1,2}$\footnotemark[1] \\
$^{1}$Department of Astronomy, University of Texas at Austin, Austin, TX 78712, USA\\
$^{2}$Department of Physics, University of Texas at Austin, Austin,
TX 78712, USA}

\date{Accepted 2009 September 25;
      Received 2009 September 24;
      in original form 2009 July 23}

\pagerange{\pageref{000}--\pageref{000}} \pubyear{2009}

\maketitle

\begin{abstract}
Observations of gamma-ray bursts by the Fermi satellite, capable of 
detecting photons in a very broad energy band: 8keV to $>$300GeV, have 
opened a new window for the study of these enigmatic explosions. It is widely 
assumed that photons of energy larger than 100 MeV are produced 
by the same source that generated lower energy photons -- at least
whenever the shape of the spectrum is a Band function. We report here 
a surprising result -- the Fermi data for a bright burst, GRB 080916C, 
unambiguously shows that the high energy photons ($\gae 10^2$MeV) were 
generated in the external shock via the synchrotron process, and the 
lower energy photons had a distinctly different source. The magnetic field 
in the region where high energy photons were produced (and also the late 
time afterglow emission region) is found to be consistent with shock 
compressed magnetic field of the circum-stellar medium. This result 
sheds light on the important question of the origin of magnetic fields 
required for gamma-ray burst afterglows. The external shock model for 
high energy radiation makes a firm prediction that can be tested with 
existing and future observations.

\end{abstract}


\section{Introduction}

The discovery of a bright gamma-ray burst (GRB), 080916C, by the
recently launched Fermi satellite is an important advance toward
our understanding of these spectacular explosions. The Large Area Telescope
(LAT) onboard the Fermi satellite can detect photons in the energy range
from 20MeV to $>$300GeV (hereafter we will call it the LAT band). 
LAT observed photons of energy up to 13 GeV from GRB 080916C where the
flux was close to the threshold of its sensitivity (Abdo et al. 2009),
and this detection suggests that the Lorentz factor of the outflow
in this explosion was $\gae10^3$ (Greiner et al. 2009).

A number of different proposals have been put forward for the generation 
of high energy photons in GRBs. For instance, one possible mechanism is 
the synchrotron process -- either electron or proton synchrotron --
e.g. Meszaros \& Rees (1994), Totani (1998), 
Zhang \& Meszaros (2001). Another possibility is inverse-Compton
scattering of lower energy photons produced in the same region 
(such as the synchrotron-self-Compton process) or of an external
origin e.g. Meszaros \& Rees (1994), Pilla \& Loeb (1998), Dermer et al. (2000), 
Sari \& Esin (2001), Wang et al. (2001a, 2001b), 
Zhang \& Meszaros (2001), Granot \& Guetta (2003), Guetta \& Granot (2003),
Piran et al. (2004),   
Beloborodov (2005), Fan et al. (2005 \& 2008), 
Fan \& Piran (2006), Wang et al. (2006), Galli \& Guetta (2008), Zou et al. (2009). 
Yet another class of
high energy photon generation mechanism is hadronic collisions and
photo-pion production e.g. Katz (1994), Derishev et al. (1999), Bahcall \& 
Meszaros (2000), Dermer \& Atoyan (2004), Razzaque \& Meszaros (2006), 
Gupta \& Zhang (2007), Fan \& Piran (2008).
Please see Fan \& Piran (2008) for a review 
of the extensive literature on high energy photon generation processes. 

In this {\it Letter} we provide multiple lines of evidence that show that high energy
photons and late time X-ray and optical afterglow emissions from 
GRB 080916C were produced via the electron synchrotron process
in the external shock;  lower energy photons ($\lae 1$MeV) had a
different origin.  In the next section we provide a summary of the
observed data for GRB 080916C. In \S3 we describe the expected high energy
emission from the external shock and compare that with the data for GRB 080916C,
and in \S4 we show that the entire optical and X-ray afterglow data for
this burst is consistent with the external shock model. Moreover, using the 
external shock parameters determined from the late afterglow data alone 
($t\gae1$day) we show that the expected emission at $>10^2$MeV during 
the prompt phase is entirely in agreement with the observed Fermi/LAT 
data (\S4). The main conclusions are summarized in \S5. 

\section{GRB 080916C: Summary of Observations}

GRB 080916C was detected by Fermi (Abdo et al. 2009) in the energy band $\sim$8keV--13GeV.
The spectrum of GRB 080916C peaked at $\sim500$ keV; the flux was 
independent of frequency below the peak, i.e. $f_{\nu} \propto \nu^{0\pm0.03}$, 
whereas above the peak a single power-law function, $f_\nu\propto 
\nu^{-1.2\pm0.03}$, extending from $\sim500$ keV to 13 GeV provided a good
fit to the data (time dependences of these quantities can be found in 
Fig. 3 of Abdo et al. 2009). The electron energy distribution index ($p$) corresponding
to this spectrum was 2.4. The LAT band photon flux
rose as $t^{6.0\pm0.5}$ during the first 4s of observations
(the time is measured starting from the first detection of photons 
in the 8keV--10MeV band), and declined as $t^{-1.2\pm0.2}$ from 4s to 1400s.
The light curve for lower energy photons on the other hand declined as $\sim
t^{-0.6}$ for the initial 55s, and subsequently it underwent a steep
decline of $t^{-3.3}$ which is often seen in the sub-MeV band of GRBs 
(Tagliaferri et al. 2005, Nousek et al. 2006) and marks the end of the emission activity of the source.
Thus, photons of energy $>10^2$MeV lagged lower energy photons by 4s, 
and that is an important discovery by Fermi. The other 
puzzling discovery is that radiation in the LAT band lasts for a 
much longer duration of time than lower energy emission.

X-ray and optical observations began about 1 day after the trigger 
time.  Optical observations allowed to determine a photometric redshift for 
this burst, $z=4.35\pm0.15$ (Greiner et al. 2009).  Using the usual 
convention, $f_{\nu}(t) \propto \nu^{-\beta} t^{-\alpha}$, the X-ray
data decayed as $\alpha_X=1.29\pm0.09$ with $\beta_X=0.49^{+0.31}_{-0.34}$, 
both values completely consistent with the shape of the optical light curve 
and its spectral energy distribution: $\alpha_O=1.40\pm0.05$ and $\beta_O=0.38\pm0.20$ 
(see Fig. 2 of Greiner et al. 2009).

Since the spectrum from 8 keV to 13 GeV had the shape of a Band
function (two power-law components smoothly joined) it has been suggested
that the observed radiation over the entire 6-decades interval in frequency
was produced by the same source (Abdo et al. 2009, Wang et al. 2009, 
Zhang \& Pe'er 2009). However, a closer analysis of the Fermi data 
shows that this possibility can be ruled out.

\section{External shock \& high energy photons}

The first evidence for two different sources of radiation -- one
dominating in the sub-MeV band and the other at $\gae10^2$MeV -- comes from 
the fact that the flux in the 50--300 keV band 
declined weakly with time ($t^{-0.6}$) during the initial 55s and then
underwent a steep decline ($t^{-3.3}$) with a distinct signature of a
short lived source of lifetime 55s\footnote{The light curve of a relativistic
source decays as $t^{-2-\beta}$ when the source is suddenly turned 
off (Kumar \& Panaitescu 2000); where $\beta$ is the spectral index which for GRB 
080916C was $\sim$1 in the 50--300 keV energy band for $t\gae55$s.}.  
This rapid decay in flux in the X-ray band has been observed 
in $\sim$60\% of all bursts detected by the Swift satellite
(Evans et al. 2009). In contrast, the source for high energy photons
-- declining as $\sim t^{-1.2}$ -- was active for at least 1400s, 
when the flux fell below the Fermi/LAT sensitivity (see Fig. 4 of Abdo et al. 2009). 
Further evidence for two 
distinct sources is provided by the detection of several other bursts 
by the Fermi satellite for which the same behavior is seen: a longer 
lasting source for high energy photons relative to sub-MeV photons
(see, e.g. Ohno et al. 2009, Cutini et al. 2009).

It is striking that the decay of the LAT light curve ($f_\nu(t)
\propto t^{-1.2\pm0.2}$) is exactly what one expects for synchrotron radiation 
from the shock heated circum-stellar medium (CSM) by the relativistic 
jet of a GRB\footnote{The shocked CSM moves with a Lorentz factor 
approximately equal to that of the GRB jet Lorentz factor. Electrons are
accelerated by the Fermi process (Blandford \& Eichler 1987) to a power-law distribution with index
$p$ such that $n(\epsilon)\propto\epsilon^{-p}$. As a result of radiative losses
the maximum electron energy is such that the synchrotron frequency
in the shocked fluid rest frame is $\sim10^2$MeV or $\sim 10^2$GeV
in the lab frame (see e.g. Cheng \& Wei 1996, Fan \& Piran 2008). 
However, this limiting synchrotron frequency depends
on the details of the electron scattering process, and it is 
likely to be higher for highly relativistic shocks.}; 
from here on we will refer to this as external shock 
or ES. We show that it is not only the time dependence of the ES emission
but also its magnitude that are the same as Fermi/LAT observations 
(with no dependence of the flux in the LAT band on unknown, and therefore
adjustable, parameters).

A number of uncertainties plague the emission calculation
from a shock-heated gas. The largest of these are the unknown strength 
of the magnetic field, and the density of the circum-stellar medium.
Fortunately, it turns out that the observed flux at a frequency
$\nu$ that is larger than all characteristic frequencies for
the shocked gas, namely the synchrotron peak and cooling frequencies,
is independent of these two highly uncertain parameters (Kumar 2000, Panaitescu \& Kumar 2000). 
Photons of energy $>$10$^2$MeV safely satisfy this frequency criterion. 
The flux in this case can be shown to be equal to 
\begin{eqnarray}
 f_{\nu} &=& {0.2 \rm mJy}\, E_{55}^{{p+2\over4}} \epsilon_e^{p-1} 
 \epsilon_{B,-2}^{{p-2\over4}} t_1^{-{3p-2\over4}} 
  \nu_8^{-{p\over 2}} (1+Y)^{-1} \nonumber \\
   && \times (1+z)^{{p+2\over4}} d_{L28}^{-2} \nonumber \\
   &=& {0.03\rm mJy}\, E_{55}^{1.1} \epsilon_e^{1.4} \epsilon_{B,-2}^{0.1}, 
\end{eqnarray}
where $\epsilon_e$ and $\epsilon_B$ are the fractions of energy of the
shocked gas in electrons and magnetic fields respectively, $t_1=t/10$s
is the time since the beginning of the explosion in the observer frame (in
units of 10s), $\nu_8$ is photon energy in units of 100MeV, 
$E_{55}\equiv E/10^{55}$erg is the scaled isotropic kinetic energy 
in the ES, $Y$ is the Compton-$Y$ parameter,
$z$ is the redshift and $d_L$ is the luminosity distance to the burst. 
The second equality in equation (1) was obtained by taking $p=2.4$, $t=4$s,
 $z=4.3$ and $d_{L28}=12.3$; $Y\lae1$ because of Klein-Nishina effects 
even though $\epsilon_e/\epsilon_B\gg1$, and furthermore, cooling o 
ES electrons by Inverse Compton scattering of prompt $\gamma$-ray photons 
can be shown to be weaker than synchrotron cooling. Note
that the flux at 100MeV is approximately proportional to $E\epsilon_e$, the
energy in electrons; it is independent of the density of the 
circum-stellar medium ($n$), and has an extremely weak dependence on 
$\epsilon_B$ which for all practical purposes can be ignored. According to
equation (1) the time dependence of the flux should be $t^{-1.3}$ 
($p=2.4$ for GRB 080916C) which is in excellent agreement with the 
observed flux decay of $t^{-1.2\pm0.2}$ in the LAT band. We note that a
good fraction of the energy of the explosion was released during the
initial 8s of the burst, and for the next 47s the
energy deposited in the external medium increased as $\sim t^{0.4}$,
and thereafter no additional energy was added to the ES. Therefore,
for $4$s~$<t<55$s the light curve decay should have been $t^{-0.9}$ due
to energy injection in ES (a slightly steeper decay -- $t^{-1.1}$ -- will 
in fact occur during this time interval due to radiative loss of ES energy),  
and for $t>55$s the decay attains the asymptotic slope of $t^{-1.3}$.
Before the deceleration time, i.e. $t<4$s, the ES light curve 
is expected to rise as $\sim t^2$ which is
significantly shallower than the observed rise of $\sim t^{6}$. This 
is a very puzzling feature that could probably shed light on the onset of the ES and 
the particle acceleration mechanism.  The observed 4s lag
for the high energy photons at the beginning of the burst is due to the
time it takes for energy transfer from GRB jet to the external shock i.e.,
the deceleration time (Sari \& Piran 1999).

For a sample of 10 well observed and studied GRB afterglows
it is found that $0.2<\epsilon_e\lae 0.8$ (Panaitescu \& Kumar 2001), and for GRB 
080916C, $E_{55}\gae 0.5$ at $t=4$s. Therefore, 
from equation (1) we find that the flux at 100 MeV from shock
heated external medium should be $\gae 2.5\mu$Jy, 
which is consistent with the observed value of 3$\mu$Jy.
It should be emphasized that this emission from the shocked external medium 
cannot be avoided. It must be present at approximately the observed 
flux value as long as electrons carry some reasonable fraction of the 
shocked gas energy (which we know is the case for GRB afterglows),
and the cooling frequency is $\lae10^2$MeV.

Does it require a coincidence for the superposition of two different 
spectra, that originated in two separate sources, to have the 
shape of a Band function? It turns out that no fine tuning or coincidence
is needed because the spectral peaks, and the flux at the peak, for ES
radiation is closely tied to the GRB jet luminosity which also regulates
the sub-MeV emission; for a very broad 
range of values for $\epsilon_B$ and $n$
the peak of $\nu f_\nu$ for the external shock emission, at the 
deceleration time of 4s, lies between $\sim$1 MeV and $10^2$MeV. 
Figure 1 shows an example of a superposition of
external shock spectrum and the sub-MeV source, and the result of a
Band function fit to it.


\begin{figure}
\centerline{\hbox{\includegraphics[width=9cm, angle=0]{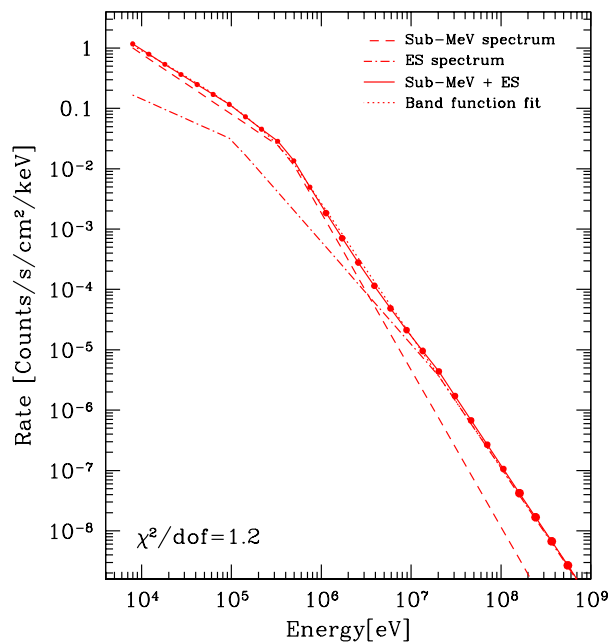}}}
\caption{ Band function fit to a superposition of external shock (ES) 
spectrum (shown as a dot-dash line) and the sub-MeV source spectrum
 (dashed line).
The superposed spectrum is shown by a solid line, and the best fit Band 
function by a dotted line ($\chi^2$/dof for the Band function fit is 1.2);
errors in the Count Rate are taken from Abdo et al. (2009), and these are equal to 
the size of filled circles. The ES spectrum is a synchrotron 
spectrum in the slow cooling regime with break frequencies 100keV
and 20MeV (values taken from the ES calculation shown in Fig. 2).
 The Sub-MeV spectrum (dashed line) peaks at 400 keV and has a slope 
of $\nu^0$ ($\nu^{-1.6}$) below (above) the peak; the choice of 
the high energy spectral index for this component is motivated 
by observations during the first 4s of the burst, when the 
emission is dominated by the sub-MeV component. If one were 
to use different break frequencies for the ES spectrum (for instance,
100keV and 70MeV), the superposition would also give an acceptable 
Band function fit.}
\label{Figure 1}
\end{figure}

We now determine the two uncertain parameters for the external shock 
mentioned previously, $\epsilon_B$ and $n$, by making use of the spectra
during the initial 55s of the burst.
The external shock emission should not dominate the observed flux 
in the 8--500 keV band since otherwise the spectrum in this band 
would be $\nu^{1/3}$ instead of the observed $\nu^0$; this means that
the flux from ES at $t=4$s between 8 keV and 500 keV should be less 
than 1 mJy (the observed flux was 2 mJy). 
This condition provides an important constraint on $\epsilon_B$ and $n$.
The flux from external shock at $\nu=100$ keV and $t=4$s is given by
(Panaitescu \& Kumar 2000, Chevalier \& Li 2000)
\begin{equation}
f_\nu = {\rm 7\,mJy}\, E_{55}^{5/6} n^{1/2} 
    \epsilon_{B,-5}^{1/3} \epsilon_e^{-2/3},
\end{equation}
For $\epsilon_e\sim 0.3$, the requirement that $f_\nu<1$mJy yields:
$n\lae 10^{-2} \epsilon_{B,-5}^{-2/3} E_{55}^{-5/3}$ cm$^{-3}$.

There is one other constraint that the external shock emission
should satisfy, and it is that the ES flux at 55s between 
50 \& 300 keV should be smaller than the observed value by at least
a factor 10 (so that the 50--300 keV light curve can decline steeply 
for $t>55$s, as observed, when the sub-MeV source turns off). 
We numerically solve for the allowed values of $\epsilon_B$ and $n$
that satisfy these two constraints, and we 
keep track of various possible ordering of characteristic frequencies.
The results are shown in Figure 2; the numerical results are consistent 
with the analytical estimate provided above. Note that there is a very
wide range of $\epsilon_B$ and $n$ allowed by the prompt data. 
Although we did not impose any constraint on $\Gamma$, its value
turns out to be $\gae2\times10^3$ -- consistent with $e^\pm$ pair opacity
argument (Abdo et al. 2009, Greiner et al. 2009). Moreover, 
Compton-$Y$-parameter at 4s is $\lae1$, even
though $\epsilon_e/\epsilon_B\gg1$, because of Klein-Nishina reduction to 
electron--photon scattering cross-section; this effect also makes the 
self-Inverse Compton scattering of ES photons undetectable by Fermi. 
 Another interesting point to note is that the entire broad range for 
$\epsilon_B$ allowed by the prompt emission data corresponds to a 
comoving-shock-frame magnetic field of $\sim100$ milli-Gauss, and that is 
of order what we expect from shock compression of a seed magnetic field in 
the circum-stellar medium of $\sim20\mu$ Gauss (pl. see Fig. 2), i.e. no 
magnetic dynamo amplification of field is 
needed behind the shock front for this burst.

\begin{figure}
\centerline{\hbox{\includegraphics[width=9cm, angle=0]{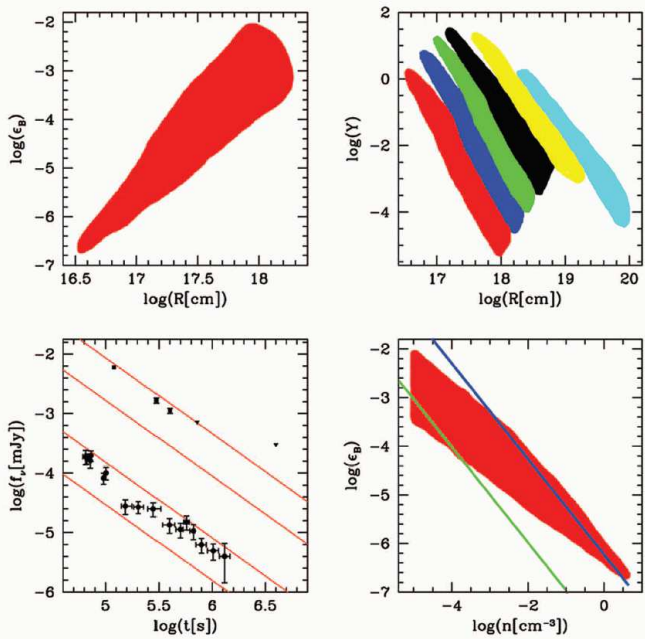}}}
\caption{ The fraction of energy of the shocked medium in the magnetic 
field, $\epsilon_B$ ({\bf top left}) at 4s (observer frame) and 
Compton-$Y$-parameter ({\bf top right}) at 4s, 15s, 50s, 150s, 1500s, 1day 
(red, blue, green, black, yellow and cyan, respectively) as a function 
of the distance from the center of the explosion to the ES front, $R$; note
that for most of the parameter space $Y\lae1$. The parameters for the
allowed solution space shown in these figures are obtained by applying
the constraints described in the text. For the allowed parameter space shown in
the upper panels we calculate the expected late time afterglow flux
 in the optical and X-ray bands as a function of observer time 
({\bf bottom left}). The
upper and lower limits for these theoretically calculated ES fluxes 
are shown as a pair of solid lines in the bottom left panel together 
with the observed flux. The optical (Greiner et al. 2009) and X-ray 
(Evans et al. 2007)
fluxes (squares and circles, respectively) are consistent with the 
theoretical expectation of the ES model (triangles are optical upper limits)
when $n$ falls off with radius 
approximately as $R^{-2}$; 
energy added to the ES during the initial 55s, and the radiative
loss of energy, was included in the calculation of late time ES flux.
$\epsilon_B$ vs. $n$ for the parameter space allowed by the prompt data 
(top panels) is displayed in the 
({\bf bottom right}) panel at 4s, and also shown is the expected $\epsilon_B$ 
for shock compression of magnetic field in the circum-stellar medium
(for CSM magnetic fields of 10 \& 70 $\mu$-Gauss -- green and blue 
lines, respectively); almost the entire $\epsilon_B$ parameter space 
allowed by the prompt $\gamma$-ray data is consistent with the
shock compressed CSM magnetic field of $\lae 70 \mu G$.
}
 \label{Figure 2}
\end{figure}

\section{Late time ($t\gae1$ day) optical \& X-ray data}

The external shock that gave rise to the high energy emission ($\gae10^2$MeV)
at early times will radiate at X-ray and optical bands at late times.
For the region of ($\epsilon_B$, $n$) parameter space allowed by the 
early time data ($t\lae 55$s) we calculate the 
X-ray and optical flux at $\gae$ 1 day after the burst, and find that these
fluxes are in good agreement with the observed values for the entire 
allowed parameter space shown in Fig. 2\footnote{The observed optical and
X-ray flux at 1 day are larger than the expected value by a factor of
$\sim 3$ for a uniform density circum-stellar medium, and these fluxes are
 smaller by about a factor $\sim 2$ when the CSM density decreases as 
$R^{-2}$. Our calculations include
the effect of energy added to the ES for the initial 55s as well as the 
radiative loss of energy. The late time afterglow data is best modeled
by a non-uniform CSM where the density falls off a little bit 
more slowly than $R^{-2}$.}. Furthermore, the observed 
spectra and light curves in these bands, $f_\nu(t)\propto \nu^{-0.5\pm 
 0.3} t^{-1.3\pm 0.1}$ (Greiner et al. 2009), are also in excellent agreement 
with theoretical expectations (the theoretically calculated values for the 
synchrotron peak and cooling frequencies at $\gae$ 1 day are $<$1eV and 
$>$1keV, respectively, for the entire parameter space allowed by the early
high energy data -- shown in the top panels of Fig. 2 -- and therefore
we expect the spectra in the optical and the X-ray bands to be $\propto
\nu^{-(p-1)/2}\propto \nu^{-0.7}$). 
The fact that the ES parameters determined from the early 
time $10^2$ MeV data provide good fit to the late time X-ray and optical 
emissions (which are well known to be from ES) lends
strong support to the interpretation that the radiation observed by 
Fermi/LAT originated in the ES. 

One could argue that the observed X-ray and optical light curves (at $t<1$day)  
could have been more complex - than a single power-law - making it difficult 
to predict the late time X-ray and optical fluxes using the ES model.  
However, optical light curves are often single power-law functions and very rarely show a
plateau (Oates et al. 2009). Moreover, a fraction of GRBs show a single power-law 
decline in their X-ray light curve and GRB 080916c could belong to this class of 
bursts (Liang et al. 2009); we note that very bright GRBs are less likely 
to contain a plateau in their X-ray light curve (Kumar, Narayan, Johnson 2008) and 
so there is a good chance that GRB 080916c - the brightest burst ever detected - 
had a simple light curve.  These arguments allow us to use the simple ES model 
to predict the X-ray and optical flux at late time and compare it with the observations. 

It is interesting to note
that this exercise works in the reverse direction as well, i.e. using the
external shock parameters determined from the optical and X-ray data for 
$t\gae1$day we calculate the flux at 10$^2$ MeV at 150s (Fig. 3; right 
hand panel), and find that to be in agreement with the observed data 
provided that we restrict $E_{55}\epsilon_e
\gae 0.1$ ($\epsilon_B$ \& $n$ can take whatever value that is allowed by 
the late time afterglow). It is also interesting to point out that
although the $\epsilon_B$--$n$ parameter space allowed by the late time
optical and X-ray data for GRB 080916C is very large (pl. see Fig. 3; 
left panel), the entire allowed range for $\epsilon_B$ by the late time
data (without making use of the early time LAT data) is consistent 
with a shock compressed circum-stellar medium magnetic field of strength
30 $\mu$-Gauss or less.  The constraints we use to obtain the ES parameters
in this reverse direction exercise are the following: (i) At 1 day the optical and X-ray frequencies
should lie between the synchrotron peak and cooling frequencies, (ii) the ES flux at 1 day should
match the observed X-ray and optical fluxes at this time, (iii) the Lorentz Factor of
the ejecta at 1 day should be $\gae 60$, so that the initial jet Lorentz Factor is $\gae 10^3$, and 
(iv) the ES flux at 150s between 50 \& 300 keV should be smaller than the observed value by at least
a factor 10 (see \S3). 

It is no small feat that
the external shock model fits the data over 10-decades in frequency
and 3-decades in time, and provides a natural explanation for a number of
puzzling features observed by Fermi during the first 10$^3$s of the burst.


\begin{figure}
\centerline{\hbox{\includegraphics[width=9cm, angle=0]{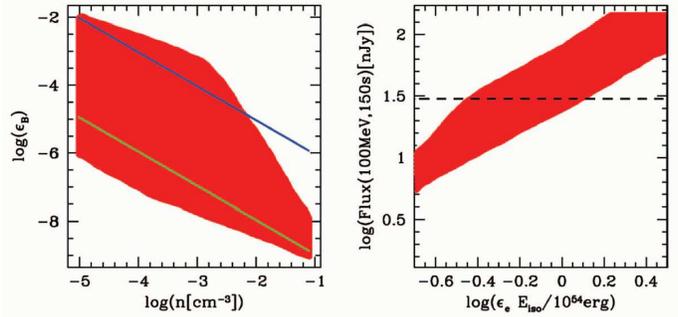}}}
\caption{ES parameters derived using the late   
($t\gae1$day) optical and X-ray afterglow data only.  $\epsilon_B$ as a function 
of $n$ ({\bf left panel}) at 150s (observer time) and the expected 
$\epsilon_B$ for shock compression of magnetic field in CSM
(for CSM magnetic fields of 1 \& 30 $\mu$-Gauss -- green and blue 
lines, respectively).  We note that the ($\epsilon_B$, $n$) space determined 
using {\bf only} the late ($t\gae1$day) optical and X-ray afterglow data 
is found to be very similar to the ($\epsilon_B$, $n$) space determined 
using {\bf only} the 100 MeV early data, and that shock compressed 
CSM field is all that is needed for the ES synchrotron emission for 
GRB 080916C. The predicted flux at 100 MeV at 150s as a function of 
$\epsilon_e E_{iso}$ using only the late ($t\gae1$day) afterglow data
({\bf right panel}). The horizontal dashed line indicates the 
observed flux by the Fermi Satellite, which was $\sim 30$nJy.   
}
\label{Figure 3}
\end{figure}


Fermi has detected a few other bursts in the high energy band.
They all seem to share similar features -- high energy photons 
lag lower energy photons initially but last for a longer
duration of time (Omodei et al. 2009). The external shock model 
provides a straightforward explanation for these ``generic'' features. 

The external shock model also makes a prediction that the fluence for 
$\nu>10^2$MeV should be proportional to $(E\epsilon_e)^{(p+2)/4} 
(\epsilon_e/t_d)^{3(p-2)/4}$ whenever the synchrotron cooling
frequency is below 100 MeV -- a condition that is easy to check from
the spectrum in the LAT band (this relation follows from eq. 1,
which provides dependence on $z$); $t_d\propto\Gamma^{-4}$ is the 
deceleration time of a GRB jet -- propagating into a wind like density 
stratified medium -- which can be taken to be the observed 
lag time for high energy photons.
This prediction can be used to confirm or disprove this model. If detectors 
are activated by flux level, rather than fluence, then they will only observe
bursts with the highest $\Gamma$ since the flux scales as $t_d^{-(3p-2)/4}
\propto \Gamma^{3p-2}$. Short duration GRBs should also satisfy 
the same scaling relation since the flux is independent of circum-stellar 
medium density.

\section{Conclusions}

We summarize the 4 main reasons that $\gae10^2$MeV photons observed by
Fermi/LAT were produced in the external shock. (1) The expected
flux from external shock at 100 MeV, at $t=4$s, is $\gae 2.5\mu$Jy (independent
of $n$ and $\epsilon_B$) and that is in good agreement with the
observed flux of 3$\mu$Jy. (2) The radiation observed by LAT lasted
for a time (1400s) much longer than the burst duration of 55s. (3) Furthermore,
the light curve decay in the LAT band, $t^{-1.2}$, is what is expected for
the external-shock emission. And so is the 4s lag for the $>10^2$ MeV 
photons. (4) The external shock parameters calculated using the initial
55s of data alone (Fig. 2), are able to explain the late time ($t \gae 1$day) 
X-ray and optical afterglow data which is widely believed to be ES emission;
as pointed out in footnote {\bf 3} the flux at late times 
depends on the density stratification of the circum-stellar medium, and
for a wide range of possible density stratification the theoretically
calculated flux lies within a factor of a few of the observed value.
Moreover, the converse is also true i.e., late time afterglow data
extrapolated back to 150s (and also to 4s) matches the observed flux at 
$\gae 10^2$MeV. 

The fact that the $\gae10^2$MeV light curve rises as $\sim t^6$ ($t<4s$) is 
puzzling, as mentioned before.  There is another 
burst (GRB 061007) that also displays an extremely rapid rise of its optical light curve
at early times (Rykoff et al. 2009) and its isotropic equivalent energy release 
is also very high (one of the highest ever recorded so far).  This suggests that 
the fast rise might be related to the particle acceleration mechanism at the onset of the
ES, but more theoretical work is needed to determine the cause of this rapid rise.

The Fermi burst (GRB 080916C) sheds a surprising light on the question
of the origin of magnetic fields in external shocks. Magnetic fields
in the source inferred from the early LAT and GBM data ($t\lae55$s) -- 
and independently calculated from the late afterglow data ($t\gae1$day) by 
itself -- are entirely consistent with a $\sim20\mu$-Gauss circum-stellar 
field compressed by the external shock, i.e. no extra field amplification is 
needed for the observed radiation (this possibility was
investigated in Granot \& K\"onigl 2003). GRB 080916C was the brightest 
burst to date, and if no magnetic dynamo is needed for the external
shock synchrotron emission for 
this burst then we suspect that this result is likely to be
applicable to other GRB afterglows as well.

\section*{Acknowledgments}
RBD dedicates this work to his wife Jessa Barniol 
and his uncle Jose Barniol.  RBD also thanks Rongfeng Shen for very 
useful discussions. We are very grateful to Dale Frail and Jonathan Granot
for valuable comments on this manuscript.  This work has been funded in part by 
NSF grant ast-0909110.  This work made use of data 
supplied by the UK Swift Science Data Centre at the University of Leicester.



\end{document}